# Jupiter's North Equatorial Belt and Jet:
# I. Cyclic expansions and planetary waves

John H. Rogers

*A report of the Jupiter Section, using data from the JUPOS team*

___________________________________________________________________

## Summary

This article presents a synopsis of the activity in Jupiter's North Equatorial Belt (NEB) from 1986 to 2010, and of the speeds of dark formations on its south edge and bright streaks ('rifts') in its interior.  In particular I discuss NEB expansion events (NEEs), which took place every 3-5 years during this time, and how the various features of the NEB are involved in them.

I present evidence that the NEE affects not just the northern edge, but the whole width of the belt.  It begins with an outbreak of a bright rift that is more northerly and slower-moving than usual; this is often involved with the first ejection of dark material northwards into the N. Tropical Zone, but typically the rift also expands southwards across the width of the NEB.  NEBs dark formations are usually affected, as they are during individual interactions with rifts at other times; they may be disrupted, or intensified, and they usually undergo deceleration.  The expansion of the dark NEB to the north occurs concurrently, and is followed by the appearance of new dark 'barges' and white ovals flanking the NEBn jet.

The speed of the NEBs dark formations varies with their mean spacing, consistent with the prevailing hypothesis that they are planetary Rossby waves.  In most apparitions since 2000 we have also detected smaller, faster features (~120 m/s). I propose that these represent waves of the same type, but with higher frequency, and that their speed is slightly less than the true wind speed at cloud-top level under normal conditions.

___________________________________________________________________

## 1. Introduction

### *1.1. The North Equatorial Belt (NEB)*

Jupiter's North Equatorial Belt (NEB) is one of the broadest and darkest belts on the planet, and is almost always a scene of notable weather formations and activity **(Figs.1&2)** [Refs.1-3]. It is nominally bounded by the retrograde jet at 17ºN on its northern edge (NEBn), and the very fast prograde jet at 7ºN on its southern edge (NEBs), although the visibly dark belt does not always respect these limits.  There is a continuous gradient of wind speed across the NEB, as shown by zonal wind profiles (ZWPs) from spacecraft, from the NEBs jet on the south edge (close to System I) to the North Tropical Current which governs the barges and AWOs on the north edge (close to System II) **(Fig.3)**.

The northern edge usually shows irregularities which often include prominent oval circulations: dark brown cyclonic ovals ('barges') at 15-16ºN, anticyclonic white ovals (AWOs) at 18-20ºN, and sometimes anticyclonic dark ovals (ADSs) at ~19ºN which may be grey or brown or reddish. The latitudes of barges and AWOs vary slightly according to their drift rate, but do not depend on whether they are surrounded by dark 'belt' or bright 'zone' material.

Within the belt there are usually white 'rifts', i.e. oblique, turbulent streaks which are cyclonic convective regions. They are often initiated or renewed by the appearance of brilliant white spots which are thought to be convective plumes arising by moist convection from the deep water-cloud layer. They are almost certainly thunderstorms, as observed in the NEB rifted region by Voyager [Ref.4 & references therein] and Galileo [Ref.5], and in the SEB by Galileo [Refs.5&6], and in similar cyclonic turbulent regions elsewhere in the planet by Voyager and Cassini [e.g. Ref.7]. The bright spots spread out obliquely as they are sheared by the wind gradient. They often expand into, or form within, more extensive rifted regions, which can last for months or even years.

On the NEB southern edge, there are large dark formations which are among the most conspicuous features on the planet, as described below. They are often associated with bright white features, which are called 'plumes', as their appearance suggests cloud streaming vertically and/or horizontally.

### *1.2. NEB dark formations (NEDFs)*

From their visual appearance, the NEBs dark formations are usually called 'projections' (from the dark NEB into the bright EZ); some that are longer but less projecting are called 'plateaux'; and long, grey streaks curving from them into the EZ are called 'festoons'. The dark formations are very bright (warm) at thermal infrared wavelengths near 5 microns, and are therefore also called 'hot spots'. 'Projections' is not an ideal term, as they can adopt a variety of shapes, and they are actually distinct weather systems, quite different from the adjacent belt. Their dark grey or blue-grey colour is different from the brown of the NEB, and they are uniquely conspicuous in the thermal infrared. Here we refer to them as NEBs dark formations (NEDFs).

The NEDFs comprise clearings in all the major cloud layers [Refs.8&9]. There are typically ~10-12 NEDFs spaced fairly evenly around the circumference [Ref.1, pp.132-145]. Their lifetimes range from months to years. They have drift rates close to System I (equivalent to wind speed $u$ = 105.5 m/s in System III), commonly in the range from ~100 to 110 m/s: the average from 1891-1991 was 107 (±2) m/s [Ref.1].

There is considerable evidence that they represent trapped Rossby waves. These are horizontal and (at this latitude) vertical meanderings of a jet, governed by the variation in the Coriolis effect with latitude, and they could be confined between the NEBs jet and the equator [Refs.10-15]. The true wind speed would be much faster than the phase speed of the Rossby waves (observed drift rate of the NEDFs), and this was verified by the Galileo Probe, which showed a speed of ~170 m/s below cloud-top level [Refs.16&17]. Moreover, the phase speed should vary according to the wavelength of the Rossby waves (spacing of the NEDFs). Evidence for such a correlation has been reported [Refs.12&18] (those results broadly agree with our data for recent years), and we have confirmed and extended this correlation in some recent reports [Refs.19 & 20, and see below]. Detailed study of the NEDF dynamics in Cassini movies [Ref.21] further supports the Rossby wave model. The Rossby wave model can also describe the South Equatorial Disturbance, a large solitary wave that is sometimes present on the SEB (Fig.1) [Refs.10,15,22,23].

*1.3. NEB Expansion Events (NEEs) and Revivals*

For most of the last century, the NEB has always been a major dark belt, not subject to the large-scale fadings and revivals for which the SEB is famous. However, in earlier decades, it did undergo drastic narrowings and vigorous revivals comparable to those of the SEB. Some of these NEB Revivals attracted attention at the time [Ref.24]; subsequent review of the literature showed that they occurred every 3 years from 1893 to 1915 [Refs.1&3]. No more NEB Revivals occurred after 1915, except for one in 1926; but more modest narrowings and broadenings were recorded every 3-5 years from 1918 to 1935.

These 'NEB expansion events' or 'broadening events' (NEEs) were first described in [Ref.1, pp.126-130]. Like NEB Revivals (but on a lesser scale), they consisted of broadening of the dark belt to the north (to ~20-21ºN), usually followed within a year or so by reddening of the expanded NEB, and by appearance of an array of barges and AWOs within the northern edge. NEB rift activity might accompany the broadening or occur later, but the visual records did not show any consistent relationship [Refs.1&3]. *[Footnote 1]*

> *Footnote 1*: Major features of NEEs seem to have been noticed by A.P. Lenham [Ref.25], although he did not describe them as a coherent process. He plotted the the width of the NEB, and an estimate of its redness, and the mean speed of the NEDFs, from 1914-1943. He reported a rough periodicity of 3-6 years in the width, and also noted probable correlation between large width, and redness, and slow drift rate [long rotation period]. We now know that the expanded NEB tends to have these properties within a year after a NEE starts (see below re speeds of NEDFs).

NEEs occurred cyclically up to 1935, and continued into the 1950s though with less regularity and/or less complete documentation [Ref.1]. However, they did not occur at all after the 1960s -- until 1987/88. The unexpected expansion of the belt in 1988 [Ref.26] was the first in a series of such cycles, again at intervals of 3-5 years, which have continued to the time of writing in 2016 **(Figs.2 & 4)**.

This article covers the years from 1986 to 2010, in which there were six NEEs. It ends prior to the great NEB fade and revival in 2011-12, which was an exaggerated version of a NEE that had not been seen for nearly a century; this cycle will be described in Papers II and III.

This time-span also covers the years of hi-res amateur imaging, which started with hi-res Kodak TP2415 film around 1985, and developed further with CCD imaging in the early 1990s, and then with webcam imaging in the early 2000s. The analysis has also been transformed, from 1998 onwards, by the JUPOS project, which systematically measures the positions of spots on images and generates detailed, comprehensive charts of their motions. Therefore, the recent activity in the NEB has been defined with far better spatial and temporal resolution than was possible in the era of visual observations.

This article will summarise the development of the six NEEs from 1987 to 2010, and the drift rates of the NEB rifts and the NEDFs during those years. We suggest that NEEs are associated with distinctive drift rates for NEB rifts and for NEDFs, and discuss whether these are respectively causes and consequences of the NEE.

The results are all derived from BAA/JUPOS analysis of amateur observations, unless otherwise stated. Most of the data have already been published in this Journal or posted in reports on the BAA Jupiter Section web site, which can be consulted for more details. The more important reports are referenced herein, and a complete list of these has been posted at:

https://www.britastro.org/node/8241.  We have not yet completed our analysis for some of these years, but preliminary data give an adequate overview.  In general, data are from the following sources.
>>1986-1992 and 1995-2002:  Full reports published in the Journal.
>>1993, 1994: Interim report [Ref.27] & unpublished BAA reports (M. Foulkes & JHR).
>>2002-2004:  Unpublished BAA/JUPOS analysis.
>>2004-2011:  BAA/JUPOS reports published on the Section web site: final reports for 2005-2007 [Refs.36&20] and interim reports for all apparitions.

## 2.  NEB Expansion Events (NEEs)

A NEE [Ref.1, pp.126-130] is characterised by broadening of the NEBn edge northwards into the N. Tropical Zone (NTropZ).  The event may start at a single longitude and spread around the planet, or it may start indistinctly with irregular extension northwards from the NEB at various longitudes.  (We describe this in visual terms as 'emission of dark material' although in reality it may be propagation of cloud-free conditions by breakup or subsidence or evaporation of clouds.)  Often it starts with a 'NEB outbreak' (NEBO), i.e. the sudden appearance of an anticyclonic dark spot (ADS) in the NTropZ in conjunction with a new or expanding rift in the northern NEB (as described more fully in Section 2.3).  **Figure 5** illustrates some of these NEEs, especially NEBOs; the diagram in **Fig.5A** is typical.

The NEE is typically followed, about a year after it has begun, by reddening of the belt, and by appearance of an array of cyclonic dark ovals (barges) and anticyclonic white ovals (AWOs) in the expanded northern NEB.  NEEs occurred in 1987-88, 1993, 1996, 2000, 2004, and 2009; then there was a full NEB Revival in 2012.

*2.1.  Chronicle of NEEs*

*(In this section, drift rates are given as DL2 in degrees per month (30 days) in System II longitude.)*

**1988-89  (NEE starting in 1988 Dec.)** *(from BAA reports in JBAA):*
The NEE apparently began in 1987 Dec. with focal disturbance (probably a NEBO) including a bright northerly rift and ADS.  However the disturbance was not characterised thereafter, as solar conjunction approached.  When the planet reappeared in summer **1988**, the NEB had broadened impressively, and showed a pattern of small-scale mottlings without large rifts – possibly a relic of disturbance which was now active on a small scale. There was just one AWO with a barge in August; more of both developed in Oct.-Dec.  Then the NEBn began to fade again in early 1989.

**1993  (focal NEE starting in 1993 March)** *(from Ref.27 & BAA reports, unpublished)* **(Fig.5A):**
In **1993**, several barges and AWOs were present before the NEE. The NEE began with a NEB outbreak (NEBO) on March 19-22, in which a slow-moving rift appeared suddenly and spectacularly, along with an ADS, adjacent to an AWO.  The ADS itself and dusky streak preceding (p.) it were prograding; the streak initiated a wave of transient patchy darkening of the NTropZ  at DL2 ~ -60 to -90 deg/month, and the ADS formed the p. end of a more persistent darkened (expanded) sector with DL2 = -35 deg/month.  The following (f.) end of the original rift was a bright white spot until May, accelerating from -30 to  -51 deg/month, and probably engendered a second eruption in late April at the original site. At other longitudes, dusky streaks streaming off NEBn dark spots contributed to the NEE.  By August, the NEB was broadened into the NTropZ almost everywhere.
By **1994** March, there were 3 AWOs and 2 barges; other barges developed up to July. The NEBn was fading again from mid-1994.

**1996  (focal NEE starting in 1996 April)** *(from BAA reports in JBAA)*: **(Fig.5B):**
**In 1996**, the NEE started with an ADS first recorded on April 6, adjacent to a very bright AWO.  A striking spotty expanded sector spread p. from there at ~-70 deg/month, though "not quite as flamboyant as in 1993". There was also irregular northwards expansion of the other longitudes of the NEB, which was completely expanded by late June.
**In 1997**, a circum-global array of barges and AWOs was present from the earliest images in April-May. 1997 Nov., the NEBn was receding again.

**1999-2000  (irregular NEE developing gradually in 2000)** *(from BAA reports in JBAA)*.
Unusually, 6  barges and 2 AWOs were already present before and during the NEE.  This NEE developed very slowly and irregularly from summer 1999 to late 2000.
**In 1999/2000**, a notable ADS ('Little Brown Spot') appeared in 1999 July, and from August onwards some dusky spots and streaks were spreading northwards from NEBn projections and at 20ºN, but with no evident pattern, and this darkening appeared to be abortive.  In 2000 Jan-Feb., the dusky material largely faded away, though the NEBn barges and ADS remained.  General broadening did not develop before solar conjunction in 2000.  **By 2000 July**, the gradual expansion event was well under way.  It was complete in one 90º sector; the next 70º sector quickly darkened thereafter; and the remaining sector showed an irregular NEBn, with dark projections, dusky streaks, and yellowish shading, all of which developed to complete the darkening by the end of 2000.
**In 2000/01,** the NEB was rich brown, possibly redder than usual, and **in 2001/02** it was notably reddish, while the northern EZ had weak yellowish shading.  There were more barges and AWOs than before, forming a typical array. The NEBn brightened again in spring 2002.

**2004  (NEE starting in early 2004)** *(from unpublished JUPOS data & bulletins for 2002-2004, recently re-posted for 2003/04 [Ref.30], and final BAA/JUPOS reports posted on-line for 2005-2007 [Refs.36 & 20])*. **(Fig.5C):**
**In early 2004**, there were three successive NEBOs in early Feb. (adjacent to white spot Z), late Feb., and early April (adjacent to the only remaining barge).  Each NEBO created a dark brown ADS in the NTropZ just after the f. end of a slowly-moving rifted region passed by.  Only the third  initiated general broadening, with  more spots appearing f. it to broaden that sector by the start of May.  Broadening was also starting at other longitudes in May as the apparition ended.
**In 2004/05,** broadening was complete when the new apparition started in 2004 Nov., and new barges and AWOs were developing; they continued to do so up to 2005 April (along a track with DL2 ~ -84 deg/month).  Slow rifts persisted until three years after the start of the NEB expansion event – and the NEB was still broad until this time. But after **2007** April, the NEBn edge began to recede.

**2009  (focal NEE starting in 2009 May)** *(from interim BAA/JUPOS reports posted on-line)* **(Fig.5D):**
**In 2009**, the NEE began with a NEBO on May 28-31 when an impressive, slow-moving rift passed a large bulge which marked the former location of a barge; a new bright cloud erupted northwards from it through the bulge, and generated a very dark grey ADS.  Two more ADSs later appeared from the same point, also prograding. Each ADS prograded rapidly at first but then drifted south again and halted or reversed its drift; the first recirculated retrograding on NEBn and then was lost.  By August, a bright AWO had developed f.  two remaining ADSs, while WSZ converged on the group from the f. side; these 4 anticyclonic circulations, separated by 4 NEBn bulges containing barges, formed a spectacular wave-like pattern until Oct.
Meanwhile, a second NEBO occurred elsewhere in August, exactly like the first.  Other spots developed in Sep.  By the end of the year, the broadening event was proceeding all round the planet, partly by small dark spots or streaks extending into the NTropZ, and partly by general yellow-brown shading around them.  Meanwhile the rift which had triggered the NEBOs expanded until the whole NEB was turbulent.
**In 2010,** the event was complete. Several new barges and AWOs already existed by 2010 April, and two more barges developed in June-July.  **In 2011/12**, the NEBn edge started to recede.

## 2.2. Involvement of NEB rifts

*(In this section, drift rates are quoted in degrees per day in System II. This section is adapted from Ref.29.)*

There is often circumstantial evidence that NEB rifts are involved with a NEE or Revival [Ref.3], but the relationship has not been clear. The event often begins with a bright rift either appearing anew, or projecting a streamer northwards which induces a dark protrusion into the NTropZ (a NEBO). Sometimes there is vigorous rift activity during the expansion event, and this is probably responsible for the effects on the NEBs dark formations (see below). However, until now it has not been possible to identify anything special about NEB rift activity in relation to NEB expansion events. Rifts are almost always present, and there is no obvious tendency for more or larger rifts during an expansion event. It would be helpful if we could quantitate the activity in NEB rifts, but this is not yet possible [Ref.29].

Review of the historical record up to 1990 found that most drift rates for rifts fell into two groups, designated *'North Intermediate Current' (NIC)* (DL2 = -2.7 to -5.0 deg/day, average -3.9 deg/day) and *'Fast North Tropical Current'* (up to -2.0 deg/day) [Ref.1, pp.123-125]. The 'fast NTropC' group mainly consisted of long-lived rifted regions on the NEBn from 1973-74 (viewed by Pioneer 10 and 11) and from 1977-82 (viewed by Voyager 1 and 2), with DL2 = -0.6 to -1.2 deg/day, within which individual white spots were arising and moving much faster with NIC speeds. But there were also several records from earlier years of visual observations in the 'fast NTropC' range, to which we will return in the Discussion. In this article, because we are considering them in comparison with other rifts, I will refer to this range as *'slow NIC'* rather than 'fast NTropC'.

A survey of observations of NEB rifts from 1986 to 2011 [Ref.29] showed that the speeds in these years largely fall into two ranges: 'fast' with DL2 = -2.9 to ~-5 deg/day, and 'slow' with DL2 = -1.0 to -2.8 deg/day [Fig.6]. Thus the two speed ranges described historically are still valid, and significant for distinguishing two types of rift. Most rifts, and white spots within them, indeed move with normal NIC speeds (hereinafter, 'fast rifts'). But rifts involved in NEEs are mostly more slow-moving than usual, in the slow NIC range ('slow rifts'). Slow rifts are observed during and sometimes after an expansion event, but are absent at other times when the NEB is neither expanding nor expanded. Although we have described the slow rifts in terms of their speed, presumably their decisive property is their northerly latitude (13-14ºN), of which the slow speed is a consequence. We have not identified any other visible characteristic to distinguish them from the faster rifts.

From this survey, the following general conclusions emerged about the behaviour of rifts in relation to each of the 5 adequately observed NEEs (omitting 1996).

1) Fast rifts or white spots (moving with the normal NIC) can be present at any time, as even slow rifts generate internal fast spots and can accelerate up to fast speed. Fast rifts are usually, perhaps always, present when the NEE begins. But substantial fast rifts disappear as the expansion event proceeds.

2) Slow rifts (moving with slow NIC; always more northerly) are not present except in relation to NEEs as described below.

3) Slow rifts always appear at the onset of the NEE, though usually not before.  A focal NEE can be initiated at the f. end of a fast rift, which induces an 'eruption'(NEBO)  in which a slow NEB rift and a NTropZ dark spot are created simultaneously, initiating the expansion (1993, 2004) (& possibly 1987-88 and 1996; insufficient data); whereas in 2009 the (moderately) slow rift arose first and then induced the NEBO.  A slow rift also appeared in the early stages of the prolonged, chaotic NEE of 1999-2000.  Sometimes it appears adjacent to a barge, if any remain (1999, 2009).

4) During the NEE, rift activity typically evolves to smaller scales (as it also does during SEB Revivals), so within about one year there may be only small slow rifts (1988) or no rifts at all (1994, 1997).  Alternatively, slow rifts may continue to appear for just over a year (1988, 2001, 2010) or up to 3 years (2004-2007), while the NEB remains broadened.

5) Slow rifts disappear and fast rifts reappear at about the time the NEBn starts to fade or recede again. (Exceptions were in 2002, when the slow rifted region remained trapped by WSZ, and in 2011, when all rifts disappeared in the run-up to the NEB Revival.)

*2.3.  Inception and progress of the NEBn expansion*

*NEB outbreaks (NEBOs) and anticyclonic dark spots (ADSs):*
A NEBO is a notable event which often defines the start of a NEE: the sudden appearance of one or more ADSs, near the Nf. end of an active rift in the NEB.  (Sometimes there are more than one NEBO, before or after the NEB expansion begins, but usually a NEBO provides the focus from which the NEE proceeds.)  Sometimes the rift undergoes a bright outburst, and/or emits a bright streamer rapidly northwards through a brown bulge on the NEBn, which then extends north to form the new ADS at 19ºN [e.g. Fig.5].  This has always occurred adjacent to a bay (AWO) and/or bulge (probably containing a barge), i.e. a pre-existing circulation on NEBn.  Even if no barge is visible, it is possible that the event always involves an inconspicuous remnant of one interacting with the rift to accelerate the white clouds northwards, as can be seen happening in more normal times in the Cassini movies and in some of our recent observations. [*Footnote 2*]

> *Footnote 2:*  Even in earlier years when no NEEs occurred, there were precedents for NEBOs where an ADS is created near the f. end of a rifted region.
> (i) R.B. Minton reported such events in hi-res photos in 1973 and 1974.  In the 1973 event, he suggested that "a large eddy in the NEBZ may have contributed to the rapid formation of LRS3 in the adjacent NTropZ." [Ref.32a].  In the 1974 event, he reported that a dark spot moved from 16.3ºN to 18.7ºN, just like the recent NEBOs [Ref.32b].
> (ii) In the Voyager images [Ref. 1 pp.120-123], retrograding 'waves' from the NEB rifted region began to recirculate in the NTropZ when they encountered a barge, and thus formed a little brown oval in the zone at 20ºN.  These phenomena were clearly similar to the recent NEBOs.

The ADS, at ~19ºN, may be either brown ('Little Brown Spot', sometimes quite reddish) or dark grey; and sometimes it is a ring with a central light spot. However, it is not methane-bright, and may be notably methane-dark, so it is not a Little Red Spot, which would have high-altitude, methane-bright cloud cover.  Usually the ADS is initially prograding (DL2 has ranged from -3 to -62 deg/month), but the motion may change drastically.  Often the ADSs oscillate in longitude, as recorded in 1999, 2004 (two), and 2009.  We have not completed latitude analysis for these spots, but it is likely that most if not all such features obey the usual zonal wind profile.

I have speculated [Ref.31] that ADSs in various domains on Jupiter, including anticyclonic vortices on jets, are shallower than other classes of ovals, which could explain why they do not appear to be not stably constrained to a characteristic latitude or speed.  This extends a previous conjecture [Ref.1, p.260 & p.270] that the small dark anticyclonic spots on jets are shallow, whereas stable ovals and barges extend much deeper.  On this model, ADSs including jetstream spots are shallow vortices, comprising little air mass, and fully entrained by the jets and  zonal wind gradients that are observed at the cloud-tops -- unlike AWOs and other ovals whose motion is less sensitive to the zonal wind profile (ZWP).  This model could explain how large ADSs can be formed by clouds from 'rifts' ejected directly across the NEBn retrograding jet. These events are easier to understand if they are entirely superficial, with the retrograding jet proceeding undisturbed underneath.

The NEBn usually coincides approximately with the retrograde NEBn jet at 17.0 [±0.4] ºN, with peak DL2 ranging from +25 to +49 deg/mth, according to spacecraft data, although it is rarely detected in ground-based observations. Could NEEs involve destabilisation of the jet?  There was some evidence for this in 1993 and 2004, when chains of small dark spots were observed f. the NEBOs (though we did not observe full jet speeds), and in 2009, when a short-lived feature did retrograde past white spot Z.  However, we have detected similar phenomena at other times [*Footnote 3*].  So there is no strong evidence that the NEBn jet plays an active role in the NEE.

> <u>Footnote 3:</u>  We often detect short-lived, modestly retrograding spots which are influenced by the jet, but we only rarely detect full jet speed, and that has been mostly for spots f. WSZ, which evidently does perturb the jet.  These f. WSZ were in 1999 (DL2 = +32 deg/month) and 2005 and 2006 (DL2 = +35 to +45) and 2009 (DL2 = +20 and +39]).

*Progress of the expansion:*
The actual NEB expansion (darkening of the adjacent NTropZ) may occur in various ways, which may occur in combination.
(1)  Propagation of an expanded sector to lower longitudes (while the f. end remains near the original L2).  This was obvious in 1993 and 1996, but even then it was not uniform nor steady. (In 1993, there was a transient wave of patchy darkening of the NTropZ with DL2 ~ -60 to -90 deg/month, but definitive darkening then spread at DL2 = -35 deg/month.   In 1996, the leading edge had DL2 ~ -70 (±9) deg/month.  In both years, there was also irregular northwards expansion elsewhere.)  In other years, the progression was not so organised (2004, 2009), or was not observed because of solar conjunction (1988, 2000).

(2)  Irregular emergence of dark material from the NEBn.  Often this appears to include streaming Np. from NEBn bulges. Various streaks and patches may appear, sometimes including grey streaks prograding rapidly at 20ºN, DL2 = -31 to -40 (1993, 1999, 2009), which prefigure the eventual new NEBn.

(3) Gradual darkening of sectors of NTropZ, typically starting with a yellowish colour.

In two focal NEEs, the expansion was essentially complete around the planet in 5 months (1993) and 3 months (1996) respectively.  Likewise it took no more than 7 months (including solar conjunction) in 1988 and 2004.  In 2000, the time of onset could not be defined, but the process seems to have taken much longer.

*Latitudes:*
Latitudes of the NEBn are shown in **Fig.4**; the values after each NEE are listed in **Table 1.** The typical latitudes, compared with NEB Revivals and NEEs in previous eras [ref.1], are as follows:

|           | Before | After |
|-----------|--------|-------|
| 1890-1916 | 11-12  | 19-21 |
| 1917-1943 | 13-14  | 20-21 |
| 1986-2010 | 16-17  | 20-21 |

Our measurements show that expansion in latitude sometimes continues even after it is apparently complete in longitude: the typical NEBn latitude is 20ºN immediately after completion, but 21ºN in the following apparition.

*Upper haze layer:*
Images in the methane absorption band at 0.89 microns show haze lying just above the main cloud deck.  Normally there are methane-dark belts approximately coinciding with the visible dark belts; indeed they tend to be more stable than the visible belts, possibly because the upper haze is constrained more by the fixed pattern of jets than by meteorological changes in the cloud layers. During the NEE in 2000-01-02, we noted [Refs.19&33] that the NEB did not broaden northwards in methane images; likewise it did not broaden in 2004 (Fig.4).  However it did broaden in 2009, though not quite as much as in visible light, and again in 2012 (Fig.4), suggesting that the NEE can sometimes disrupt the overlying haze.

Another striking feature in 2000-01 was a regular pattern of methane-dark waves overlying the newly expanded NEB, which also coincided with thermal waves detected by the Cassini spacecraft [Refs.33-35].  Otherwise, such methane-dark waves have only been detected in a limited sector in the 2009 NEE, until now.  At the time of writing in 2016, a similar wave pattern has reappeared during a new NEE.  Whether this is a general phenomenon is a question for further collaboration with professional astronomers.

### 2.4. Later sequels

*Origin of barges and AWOs:*
Most or all barges and AWOs usually disappear before a NEE, except for WSZ.  A new set of them develops all around the planet after the NEE is complete; a total of 12-14 such circulations is typical.  The pattern can be seen from the long-term JUPOS chart from the domain, posted in *Appendix I (Supp. Online Material).*

In the best-documented cases (1988, 2004), a few new barges and AWOs have appeared as early as 7-8 months after the onset of the NEE (as did one barge in 1996); then more appeared at 10-12 months.  In other cases the timing was obscured by solar conjunction, but they developed within 12 months of the onset, and sometimes continued to appear up to 14-16 months (1994, 2010).

The origin of individual circulations is almost imperceptible, as they begin very small and inconspicuous.  It probably resembles the origin of barges that has been documented more locally in the wake of a slow rifted region (1979) or of WSZ (2005 & 2006: Ref.36). New small barges commonly form several tens of degrees f. WSZ, and in 2006 we recorded a new barge forming there by mergers of streaks retrograding in the NEBn jet.

*Reddening of the NEB:*
We have not attempted any systematic study of colour changes so far. Although there is a wealth of colour information in modern images, the colour balance and saturation in images as presented is largely arbitrary, and interpretation of it is still entirely subjective.  The images did generally show the NEB as unusually red in 1997 and in 2001/02, probably as sequels to the NEEs of 1996 and 2000. It is unclear whether there was any reddening after the other NEEs. (Conversely, the new NEBn edge is usually a narrow grey or bluish-grey band centred at 20ºN.)

*Clearing of the expanded NEBn:*
The subsequent clearing of the expanded northern NEB, by fading or by southwards recession, can begin anything from 1 to 3 years after the onset of the NEE.

## 2.5.  Overview and discussion of the NEE process

It has long been thought that NEB Revivals are comparable to SEB Revivals [Refs.1,3,24], in spite of major differences in the way they proceed; and that NEEs may be a similar process in which NEB rifts and NEDFs are also involved [Refs.1 & 3]. But until recently, it has not been possible to substantiate this conjecture.

It is notable that NEEs commonly start with the appearance of a new, brilliant white spot that becomes a rift. This enhances their similarity with comparable grand disturbances in other domains – SEB Revivals and NTB jetstream outbreaks [Ref.20].  But in the SEB and NTB, this brilliant convective plume is clearly the initiating feature of the disturbance, marking a source which remains coherent for many weeks: in the SEB the plume emerges from a stationary persistent source, and in the NTB it persists on the peak of the prograde jet.  In contrast the NEE does not always have a single source, and even if it does, the new white spot does not remain there; rather, it typically drifts with the zonal wind (i.e., is advected) and is sheared into an expanding rift by the wind gradient in the belt. It is worth considering whether the slow, northerly rifts are more powerful than the typical faster rifts, in terms of the energy released in them or their ability to affect neighbouring regions, but this idea has not yet been developed quantitatively.

So it remains to be determined whether the plume in the NEB – the new slow rift – is a cause or effect of the initiation of the NEE.   The early stages of the NEE typically include rapid protrusion of material northwards from the NEB, formation of ADS(s) in the NTropZ, as well as diffuse darkening in the NTropZ.  The formation of the new slow rift, often a sudden conspicuous event, could be just one of these induced phenomena (which itself can go on to induce others), or it could be the driving force behind them [Ref.29].  The former option would be consistent with the apparently chaotic situation in the NEB; the latter would be more comparable to the SEB Revival.

Although we still do not have a complete understanding of NEEs, much is now coming into focus.  We have identified the following components of the process, and as discussed above, the following order may – or may not – be a causal sequence.

1.  Some unknown process causes the dark NEB to shrink in latitude, usually with a 3-5-year periodicity.  This primes it for a Revival or NEE.
2.  A rift appears which is more northerly (and hence slower-moving) than usual.
3.  This triggers one or more local NEBn outbreak(s), propelling dark material into the NTropZ.

4.  The darkening of southern NTropZ (by breakup of the white cloud layer) becomes a self-propagating process, advancing to lower longitudes and/or developing diffusely; while the rift system expands across the NEB.
5.  In at least some NEEs, a regular pattern of methane-dark waves develops in the haze over the expanding or expanded NEB.
6.  Meanwhile, the rift system destabilises the NEBs jet, changing the pattern of NEDFs in various ways, and reducing their drift rate (see below).
7.  Turbulence in the rift systems evolves to smaller scales, and normal fast rifts are suppressed, although slow rifts may persist as long as the NEB is broad.
8.  An 'orange flush' may diffusely overlie the broadened NEB.
9.  After the NEBn expansion is complete, arrays of barges and AWOs develop on either side of the NEBn jet.  This may be a manifestation of a general phenomenon, that barges and AWOs are created to take up surplus energy and vorticity from the NEBn jet, especially in the wake of a rift or of WSZ, or in the energetic aftermath of a NEE [Ref.1 p.265].

**Table 1.   NEB expansion events & behaviour of NEBs projections during them.**

| Year | Start date | Type | Final NEBn lat. | NEDFs: Positive DL1? | Appearance of NEDFs: Disrupted or fading? | New or enlarged? | Reversed shape? |
|---|---|---|---|---|---|---|---|
| 1988 | 1987 Dec. | focal | 22.0 (±0.3) (1988) | Y | Y | - | Y |
| 1993 | 1993 March | focal | 20.3 (±1.2) (1994) | (y)* | (-) | (y) | - |
| 1996 | 1996 April | focal | 20.1 (1996), 21.4 (1997) | Y | Y | Y | - |
| 1999-2000 | | gradual | 20.5 (2000/01), 21.3 (2001/02) | Y | Y | - | Y |
| 2004 | 2004 April | gradual | 20.4 (±0.3) (2006) | Y | - | (y) | - |
| 2009 | 2009 May | focal | 20.7 (±0.3) (2010) | Y | - | Y | Y |

Y, yes; -, no.   [*Positive DL1 was recorded after the 1993 NEE but did not reach the threshold of DL1 = +7.5 defined in section 4.]

## 3.  Interaction of NEBs dark formations with NEB rifts and NEEs

*(In this section, drift rates are given as DL1, degrees per month (30 days) in System I.)*

It has long been known that NEB rifts can have notable effects on NEDFs as they pass [Refs.1&2].  As the rift passes, or within a few days afterwards, the NEDF may be disrupted, or enlarged, or intensified; and dark patches from them may adopt positive DL1 ('retrograding').  These effects were extensively documented from visual observations and also from Voyager imaging.  More recently, several interactions of the same type have been described in detail from Cassini imaging [Ref.21; see their Figs. 6 & 7].

We have recorded similar interactions on various occasions from hi-res amateur imaging, e.g. in 2003/04 [Ref.30] and in 2005 [Ref.36].  Here, I investigate whether they are involved in NEEs.  This is not yet a systematic survey of the forms of NEDFs – which would be a considerable challenge, in view of all the variable parameters that could be described, and the difficulty of quantitating them.  A potential relationship was apparent during the 2009 NEE, which was strikingly characterised by the sudden reappearance of the dark NEDFs after a year's absence, and by their unusually slow drift rate, and by the abnormal 'reversed' appearance of many of them (i.e. projecting at the Sp. instead of the Sf. end; e.g. 2009 in **Fig.2**) [Ref.37]. The 2009 event was observed at higher resolution than any previous example. Comparison with previous NEEs reveals that some of these phenomena do commonly occur in these events; which may be simply because NEEs always (though not exclusively) involve vigorous NEB rifts.

The behaviour of NEDFs during NEB expansion events is summarised here [from Ref.37] and in **Table 1**. (A more comprehensive summary of their appearances from 1986-2010 is in *Appendix III [Supp. Online Material]*.)

1988 (NEE started 1987 Dec.): NEDFs reappeared in late 1987 (before the NEBO), but were variable, then they largely faded and in 1988/89 were small and retrograding, generally with 'reversed' shape.

1993 (NEE started in March): In 1993, the NEBs appearance was typical with many stable projections with festoons throughout the apparition, mostly near-stationary in L1. After the NEE started, some were disrupted or swollen by rifts, and the swollen NEBs projs. were modestly retrograding. In 1994, the appearance of NEDFs was still typical, though faint at some longitudes, but all had modestly retrograding drifts (mean DL1 = +5 deg/month).

1996 (NEE started in April): After the NEE started, disruption/enlargement of NEDFs was noted, and most NEDFs developed retrograding drifts (in contrast to 1995).

1999-2000: In 1999 and early 2000, NEDFs were very dark, classic projections and near-stationary; then, gradually, they largely disappeared in late 2000. All the remaining ones were retrograding; so the transition was similar to 1988/89.

2004 (started in April): There was a typical array of prominent projections, but strongly retrograding from as early as 2003 Nov. (DL1 ~ +8), and remarkably so around half the circumference from 2004 Jan-May (DL1 = +14 to +21). Individual NEDFs interacted strongly with a major rift system, but this was occurring as early as Jan. and the rift system was fast-moving. So all of this occurred before the NEE, and even before the first NEBO in Feb. A slow-moving rift system emerged from the large fast-moving one in early Feb.; this triggered a NEBO which initiated the NEE in early April, and at the same time, it strongly disturbed two NEDFs passing it [projs. m and a; **Fig.5C**]. Further dramatic interactions continued in May, though the overall appearances of NEDFs did not systematically change; there were no 'reversed' aspects. In 2005 Jan. there were just six long 'plateaux', still strongly retrograding (DL1 ~ +14 to +23), until they broke up in Feb.

2009 [Ref.37]: All large NEDFs had disappeared in mid-2008 (see 'super-fast speeds', below). In 2009 there were few features on NEBs, apart from minor transient festoons with fast speeds (DL1 ~ -20 to -30 deg/mth), up to June. But in July, as the NEE developed, the NEBs was transformed with the appearance of many retrograding spots, induced by the expanding slow-moving rift system but spreading to all longitudes. These included small transient dark spots in the NEBs (8-10ºN; DL1 ~ +30 to +40), and more impressively, large plateaux (NEDFs) which developed alongside the rift system from the start of July onwards. These NEDFs were retrograding with DL1 ~ +10 to +17, and by Sep. they had a spacing of 29º around most of the planet. Many of the dark plateaux/projections had 'reversed' tilt [e.g. **Fig.2**].

___________________________________________________________________

During each NEE, the major NEDFs showed a shift to more positive (retrograding) DL1, relative to their speeds in other years (discussed further in the next section). The changes in appearance of the NEDFs were more diverse; in different years they were disrupted, or disappeared, or enlarged. The 'reversed' shape of the projections was noted after the 1988 and 1999-2000 events, as in 2009, but not after others. So we cannot yet establish a consistent pattern of shape changes. In summary, common aspects of the NEDFs during NEEs may be as follows: (i) positive DL1 ('retrograding') (see below); (ii) change in appearance, becoming either more or less conspicuous; (iii) possibly, 'reversed' shape.

Our working hypothesis now is that vigorous rifting is indeed a central component of NEEs, and that it induces diverse changes in the NEDFs globally just like the changes that have been reported during local interactions [Refs.1,2,21]. It is not clear whether the slow-moving rift systems associated with NEEs have any special role, or if the changes to NEDFs merely reflect the necessary presence of rifts at that time. Retrograding speed seems to be most specifically associated with NEEs, but it is instructive that in 2004, the NEDFs were already retrograding and interacting with a large rift system several months before the NEE. Whether this was coincidence, or an early aspect of the NEE process, cannot be decided at present.

In 2009, the NEE and associated rifts clearly induced the reappearance of NEDFs after a year's absence [Ref.37]. Likewise, in 1893-1915, NEDFs were not normally present in this era, but new dark spots did appear on NEBs during several of the classical NEB Revivals in those years [Ref.1 pp.127-128]. (The NEDFs had positive DL1 after the 1896 and 1906 Revivals, though not after that of 1912.) The familiar typical array of large NEDFs first became established during the 1912-13 NEB Revival. A dramatic modern instance of this would occur exactly a century later, in 2012. As we will discuss in Paper II, these events suggest that NEB rifts may actually be necessary to sustain the Rossby wave pattern that is manifest as NEDFs.

## 4. NEBs dark formations and the NEBs jet

*(In this section, speeds will be quoted both as DL1 (degrees per 30 days relative to System I, corresponding to the direct measurements), and as **u** (m/s relative to System III, referenced to latitude 7.0ºN, which may be more convenient to accommodate the large range of eastward speeds). u = —— (DL1—221) x 0.47763. )*

### 4.1. Drift rates of NEDFs

First I review the drift rates of the NEDFs since 1986. This preliminary survey is summarised in **Table 2 & Fig.7.** This chart follows on from the historical chart in [Ref.1 p.144-5], which showed a very slow, fluctuating decline in speed since 1913, punctuated by six marked decelerations that lasted only a year or so. It was noted that five of the six (in 1906/07, 1953/54, 1965/66, 1975/76, and 1988/89) occurred at the season of maximum solar heating of the NEB: 1-2 years after perihelion, and within one year of the greatest northerly latitude of the sun. However the 1921 deceleration was not at this season, and other perihelia were not accompanied by decelerations. Also, it was noted that the 1906/07, 1953/54, and 1988/89 decelerations coincided with NEEs – though the others did not.

The new chart shows the mean drift was normal and stable from 1990-1995 (as were the appearances of the NEDFs), but much more variable since then (ditto). It again shows distinct decelerations, in 1988/89, 1996, 1999-2005 (especially 2004-05), and 2009-11. To some extent these support the previous correlations both with perihelia (which occurred in 1987, 1999, and 2010-11) and with NEEs (each of which coincided with a deceleration, though it was trivial in 1993, and preceded the NEE in 2004). If a mean DL1 = +7.5 deg/mth (u = +102 m/s) is taken as the dividing line, slow speeds were associated with every NEE except (marginally) 1993. However, these correlations do not explain why the most persistent deceleration started in 1998 (just before perihelion) [*Footnote 4]* and lasted until 2006 – except that it continued the overall decline seen since 1913. The mean DL1 = +10 deg/mth (u = +100.8 m/s) was much slower than for any other 7-year interval on record.

> *Footnote 4:* The drifts in 1998/99 are shown in the BAA/JUPOS chart [*Appendix IV: Supp. Online Material*]. In this apparition, as described in our BAA report [Ref.38], the NEDFs had very variable drifts and morphologies, which often seemed to be affected by passing rifts, but seemed chaotic. In retrospect, we can see that the shift from fast to slow drifts involved two concurrent sets of NEDFs with different speeds and spacings, which overlapped for several months – as in 1997 [Ref.18] and in 2007 [Ref.20] (see section 4.3).

Nothing is known of the causes of these decelerations. The speeds of NEDFs do not depend on their latitude [e.g. Ref.36]. Given that a major determinant of the NEDFs' speed is their spacing (see below), it is possible that the NEE, with its associated rifting, primarily affects the spacing, perhaps by varying the power or pulsatility of transmission of energy from the convective rift systems to the wave pattern on the NEBs jet. However, in the 2009 NEE the speed was more decelerated than the spacing would predict (see Fig.8 below), suggesting that the speed may be primarily affected.

While the NEDFs have developed slower drifts in recent years, they have also usually had more variable and irregular appearance. However, in 1999 and again in 2006, the NEDFs formed a more regular periodic array with faster speed (almost equal to System I), concurrently becoming much darker and more conspicuous, along with more general darkening of the Equatorial Zone. It is not clear what caused these apparently coordinated changes. In 2006 the EZ darkening was brownish grey, and seems to have been the first component of the global upheaval of 2006-07; but in 1999 the EZ darkening was bluish grey and was not associated with changes in other latitudes.

At other times, there is a general disappearance of NEDFs, and this too may have various causes that are not yet understood. In 2000, the number of NEDFs was reduced and a thick bright cloud layer formed over most of the EZ [Ref.33]. In 2008 and 2010-2012, the NEDFs disappeared from large sectors while the NEB as a whole was quiescent and narrowed, before and after the 2009 expansion event. The remarkable speeds that appeared in these NEDF-free sectors will be mentioned below and fully described in Paper II.

## *4.2. Fast-moving features (u ~ 120 m/s) [Table 2]*

Before 2001, speeds faster than DL1 = -20 (u = +115 m/s) were rare. Nothing on the NEBs had been reported to move faster than DL1 = -29 (u = +119.4 m/s), except for a single spot with DL1 = -50 (u = +129.4 m/s) recorded visually, and a single record from HST. However in 2001/02, the JUPOS drift chart had a novel appearance [Ref.19]. While there were only 4 or 5 large NEDFs, which appeared as long low plateaux, there were also numerous small spots moving exceptionally fast (DL1 = -14 to -36; mean DL1 = -26, u = 117.9 m/s). They were both bright spots and dark projections, typically lasting ~3-6 weeks. The dark spots were at 7.6ºN, the same latitude as the normal dark formations, and they had a fairly regular spacing of ~13º longitude [Ref.19].

Similar fast-moving small spots have been seen in most apparitions since then (see **Table 2**), with speeds ranging up to DL1 ~ -40 (u ~ +125 m/s). They included similar numbers of dark and bright spots. The dark ones, miniature projections, always had a mean latitude between 7.0-7.6 ºN. The bright ones had more diverse latitudes ranging from 5.8ºN to 8.2ºN. These features were usually too small to have been tracked by visual observers, so there is no reason to think they were a new phenomenon.

In 2008, a conspicuous bright rift that developed in southern NEB from May onwards, with drift intermediate between Systems 1 and 2, led to the breakup of all the large NEBs formations in one hemisphere as it passed them during May-June.  In its wake, in late May-June, one group of small short-lived features appeared with DL1 = -38; then in early July, more disrupted plateaux were replaced by a more extensive set of closely-spaced dark projections, which soon formed an array with DL1 ranging from -28 to -40 deg/mth (average DL1 = -35, u = +122.3; spacing 12-18º). In this sector, some of the fast-moving projections were quite large and dark.  They were also dark in methane images, just like slower-moving major projections in this apparition.  So they appeared to be normal projections apart from their speed.  However, a sample that were investigated did shrink from large to medium or small size soon after they adopted their rapid drift, so we cannot be sure that such features would have been tracked in the pre-JUPOS era.

### *4.3. Super-fast features, 2008-2012 (u ~ 140 m/s)*

In 2008 July, while large NEDFs were replaced by smaller fast projections around half the planet, the remaining large formation in another sector was replaced by even faster features, with average DL1 = -60 (u = 134 m/s). This was faster than ever observed before on the NEBs [Ref.39].
      In spring, 2009, the NEBs was still devoid of large formations and all the drifts recorded were in the fast, but not super-fast, range.  This state ended dramatically in July when a vigorous expanding rift system, associated with the ongoing NEE, apparently induced the formation of many retrograding dark spots and NEDFs on NEBs.  However, they did not last through the next year.
      By 2010 Sep., only five major NEDFs remained – long low plateaux – and these too were subsiding. As the NEDFs diminished, a few dark features appeared with DL1 = -29 to -36 deg/mth (u = 119.4 – 122.7 m/s); and then, some much faster.  In the 2011 apparition, the disappearance of the normal NEDFs resumed and the NEBs became completely taken over by super-fast speeds, which continued to accelerate, until in 2012 Jan-Feb., there was a mean DL1 = -83 deg/mth (u = 145 m/s).  These unprecedented speeds will be discussed in Paper II.

### *4.4. Discussion: speeds on the NEBs jet*

Determinations of the mean NEBs jet peak speed from spacecraft imaging are listed in **Table 3**. As explained in the footnote, the speeds from Voyager and HST represent the drifts of the NEDFs at the time, and the speed from Cassini is probably an average between the NEDFs and the small super-fast features that were unique to that data set.  The speed from New Horizons represents the small fast features that we discuss below.

**Table 2:  Our records of drifts on the NEBs/EZn, 2001-2011**
**(L) normal speeds for NEDFs; (R) fast and super-fast speeds:**

| Year | Main NEDFs No. | DL1 | Rapid spots & projections: Average/Consensus (Range) DL1 | u3 |
|---|---|---|---|---|
| 1986-88* | 11 (collapsing) | +3 to +5 | -23 (-18 to -29) (p. ends) | 116.5 (114 to 119) |
| 2001/02 | 4-5 | +12 ( ( | -26 (-14 to -36); -56 | 118 (112 to 123); 132 |
| 2002/03 | many (transient) | 0 to +8 (var.) | -23 (-14 to -45) | 116 (112 to 127) |
| 2003/04 | 11 | +8 to +21 | -18 (-12 to -35) | 114 (111 to 122) |
| 2004(late) | 6 | +19 (±4) | | |
| 2005(early) | 11 | +13 (±2) | -- | |
| 2006 | 12-13 | +1 (±4) | *-17,-28 [w.ss. in plume cores]* | |
| 2007 | 12  ( 9 ( 9 | +10 (±2.5) -3 (±3.5) | -21 (-17 to -37) -45 | 115.5 (114 to 123) 127 |
| 2008 | 3 | +4  ( ( | -35 (-28 to -40) -60 (-45 to -66) | 122 (119 to 125) 134 (127 to 137) |
| 2009 | 0 --> 10 | +10 to +17 (prelim) | -27 (-18 to -40) | 122 (118 to 125) |
| 2010 | 5 --> 6-8 | +13  ( ( +13 to +39 | -40 (-30 to -49) -68 (-57 to -78) -- | 125 (120 to 129) 138 (133 to 143) |
| 2011/12 | 0 | -- | -71 (-36 to -95) | 139.5 (123 to 151) |

___________________________________________________________________

This table lists all apparitions in which fast or super-fast speeds were observed. There is not a complete separation of fast and super-fast speeds; speeds in the range DL1 = -40 to -50 can be assigned to either class. Descriptions of the appearance of NEDFs in these years, 1986-2011, are given in *Appendix III (Suppl.Online Materials)*.

*In 1986-87, large dark plateaux on NEBs 'collapsed' as rifts passed, then disappeared, and their p. ends appeared to drift with these rapid speeds. In retrospect, these probably represented smaller rapidly-moving spots arising within the subsiding NEDFs.
___________________________________________________________________

### Table 3: Peak speed of the NEBs jet

|  | B" Graphic | u3 (m/s) | DL1 (/mth) |
|---|---|---|---|
| **From Voyager, 1979** | +8 to +5 | 103 | 5,0 |
| **From Hubble (1995-1998)** | +5 to +8 | 105,0 | 1,0 |
| **From Cassini, 2000** | 6,8 | 113,9 | -17,4 |
| **From New Horizons, 2007** | 7,5 | 112,7 | -15,2 |
| mean | 7,2 | 108,7 | -6,7 |
|  |  |  |  |
| Historical (1900-1991) | 7 | 107 | -3,0 |

*FOOTNOTES:*

*Voyagers, in 1979:* [Ref.40]: Mean $u$ = 103 m/s, the same as the NEDFs at the time. Hi-res Voyager images showed no systematic motions relative to the NEDFs [Ref.41].

*Hubble Space Telescope (HST), in 1995-1998:* [Ref.42]: Mean $u$ = 105 m/s: Individual charts for 6 dates in 1995-1998 all showed the peak in the range ~100-110 m/s, so it was always dominated by the normal NEDFs. Re-analysis [Ref.43] reached similar conclusions, although with some points encroaching into the 'fast' range.

Beebe et al. [Ref.44] obtained the same result in 1995 images, but in HST images taken on 1994 July 29, they found a peak of 150 (±10) m/s – the only detection of super-fast speeds in HST images before 2008. Not included here are data from HST in 2008, which showed a peak speed of ~131-155 m/s over large sectors [Refs.15 & 45], confirming our report [Ref.39].

*Cassini, in 2000 Oct-Dec:* Correlation analysis [Ref.34] found mean $u$ = 114 m/s, with very high scatter (±20 m/s) – probably due to a mixture of normal NEDF drifts and much faster speeds. There were only 7 NEDFs, with some long gaps between them, but the EZ was largely covered in unusually thick white clouds [Ref.33].

Also, the near-IR images revealed widespread super-fast speeds: large sectors at ~140-150 m/s, and some up to ~170 m/s [Refs.45-47 & 15]. The latter were mainly small white 'scooter clouds', possibly lying at greater depth where the Galileo Probe had revealed similar speed. Whether the super-fast speed is the true normal cloud-top wind speed will be considered in Paper II.

*New Horizons, in 2007 Jan:* [Ref.48]: Mean $u$ = 113 m/s, i.e. in the fast range – suggesting that the fast motions that we detected in limited sectors in 2007 were even more widespread at the time.

### *Significance of the fast speeds*

The fast features (~120 m/s) are probably common, but were usually undetectable until the advent of modern amateur imaging and analysis. They have been detected in most years since 2001, and probably also in 1986-88.

In most of these years, they have been seen especially (though not exclusively) where the normal large NEBs projections had broken up, sometimes due to passing rifts in the NEB. But they are sometimes superimposed on the normal formations, as we reported in 2001/02 and 2007. They may co-exist with large formations more commonly, but the JUPOS measurers are less likely to separate them when the two types are superimposed. However they seem to have been genuinely absent in years when the large NEDFs were very dark and numerous and well-organised, such as 1979 (Voyager) and 1999 and 2006.

The shape of these features resembles the NEDFs in miniature, but is not diagnostic of their nature because a variety of weather systems could adopt a similar shape when travelling along a jet. However, a further reason for believing that they are similar to NEDFs is the relationship between speed and spacing. When there is a fairly regular series of these fast features, as in 2001/02 and 2008, the speeds and spacings are roughly consistent with, and extend, the relationship established for NEDFs [Ref.19], as described below **(Fig.8)**. In that case, their speed of ~120 m/s is a lower limit for the true peak wind speed of the NEBs jet.

### *NEDFs as planetary waves (Rossby waves).*

Our records of speeds for NEDFs, and for the faster features, generally reinforce the reported correlation of speed with spacing for NEDFs [Refs.12 & 18]. Our correlation line **(Fig.8)** is not significantly different from theirs [Ref.18]. In our records for 2001/02, wide spacing for slow features extended the correlation to lower values of speed and wavenumber than previously, while the fast features extended the correlation to higher values [Ref.19]. Arregi *et al.*[Ref.18] found two separate systems in 1997 with different speeds and spacings, and re-analysis of our chart for 1998 shows the same phenomenon [*Appendix IV: Supp.Online Material*]. In those two apparitions, the 'zigzag' nature of some tracks suggested that the two co-existing wave systems were intersecting. This was shown even more persuasively in 2007, when there appeared to be two concurrent sets of NEDFs – one set of compact projections with zero or slightly negative DL1, and a set of longer, more widely spaced projections described as 'plateaux' with positive DL1 – and these behaved like intersecting wavetrains, crossing over each other and showing constructive and destructive interference [Ref.20]. **Fig.8** shows all our results for these and some other apparitions (only including apparitions where the spacing and speed is reasonably uniform for at least some NEDFs). Despite some scatter, this chart strongly confirms the previously proposed correlation, and thus supports the Rossby wave hypothesis.

Therefore it is likely that the fast features are the highest-frequency members of this class of waves. The phase speed of the highest-frequency Rossby waves should converge on the actual wind speed, so extrapolation to zero spacing should give the true wind speed of the NEBs jet. Given the scatter and non-linearity of the chart, it does not give a unique value, but it suggests that this must be at least DL1 ~ -60 deg/month (u ~ 134 m/s), and is consistent with the value of ~140 m/s inferred by Arregi *et al.*[Ref.18]. It could even be as high as ~170 m/s (as detected lower down by the Galileo Probe).

Important new evidence on the speed of the jet came in 2008-2012, as the continuing dissolution of the NEDFs led to unprecedented take-over of the NEBs by super-fast speeds up to 150 m/s. This topic will be considered in Paper II.

## Acknowledgements

The analysis is based on the work of numerous amateur observers (who are listed in our final reports for each apparition and on the JUPOS web site), and of the JUPOS team (Hans-Jörg Mettig, Gianluigi Adamoli, Michel Jacquesson, Marco Vedovato, and Grischa Hahn). I am greatly indebted to all of them for the opportunity to produce this synthesis of their work.

______________________________________________________________________

## *Figures (small copies)*

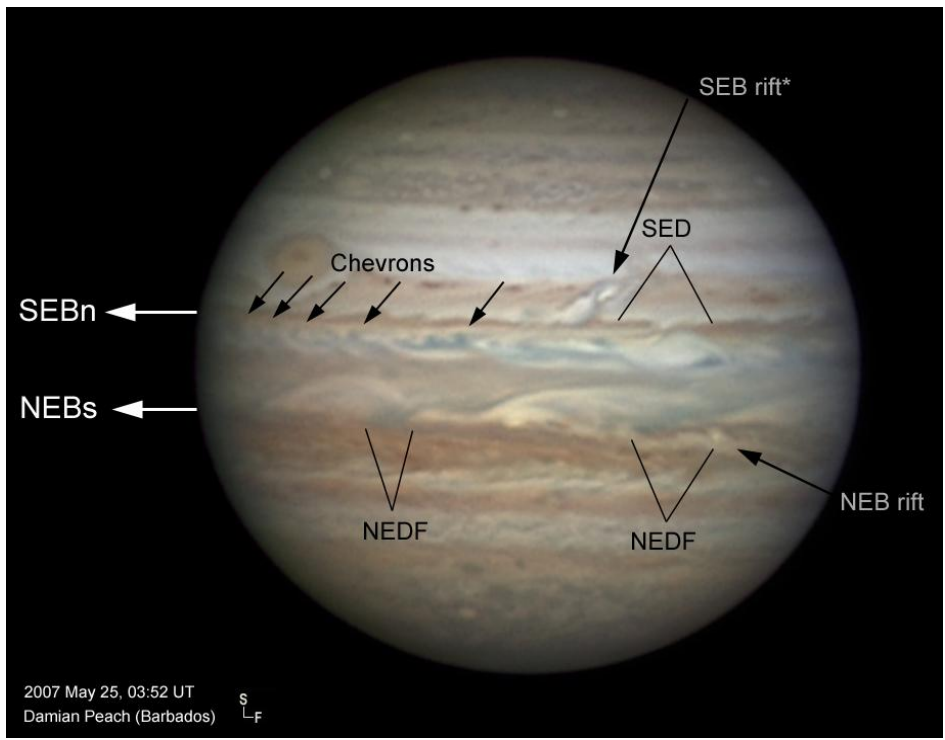

**Figure 1.** Hi-res image of the planet showing typical NEDFs, and a comparable large-scale wave structure in the southern Equatorial Zone called the South Equatorial Disturbance (SED), as well as small super-fast spots on the SEBn jet called chevrons, and rifts in both NEB and SEB. (Image taken on 2007 May 25 by Damian Peach.) *South is up in all figures.*

______________

*[Next page:]*
**Figure 2.** Alignments of map sectors from 1995-2010, showing NEEs. These all include the GRS. A similar set showing white spot Z (WSZ) is in our long-term report on it [Ref.28]; it is also marked in a few panels here, as is oval BA. Scale marks at bottom of some panels are at intervals of 30° longitude. Sources are as follows: for 1995-1999, images by I. Miyazaki, maps by H-J. Mettig (mostly published in our reports in the Journal). For 2000-2002, images by A. Cidadao and T. Akutsu, maps by H-J. Mettig. For 2004, images by D.C. Parker, map by D. Peach. For 2003, 2005, 2006, 2007, and 2010, images and map by D. Peach. For 2008, images by A. Wesley and M. Salway, map by M. Vedovato. For 2009, images by T. Barry, C. Go & T. Akutsu, map by M. Vedovato. Intensities and colours have been arbitrarily adjusted so should not be used for comparisons. *South is up in all figures.*

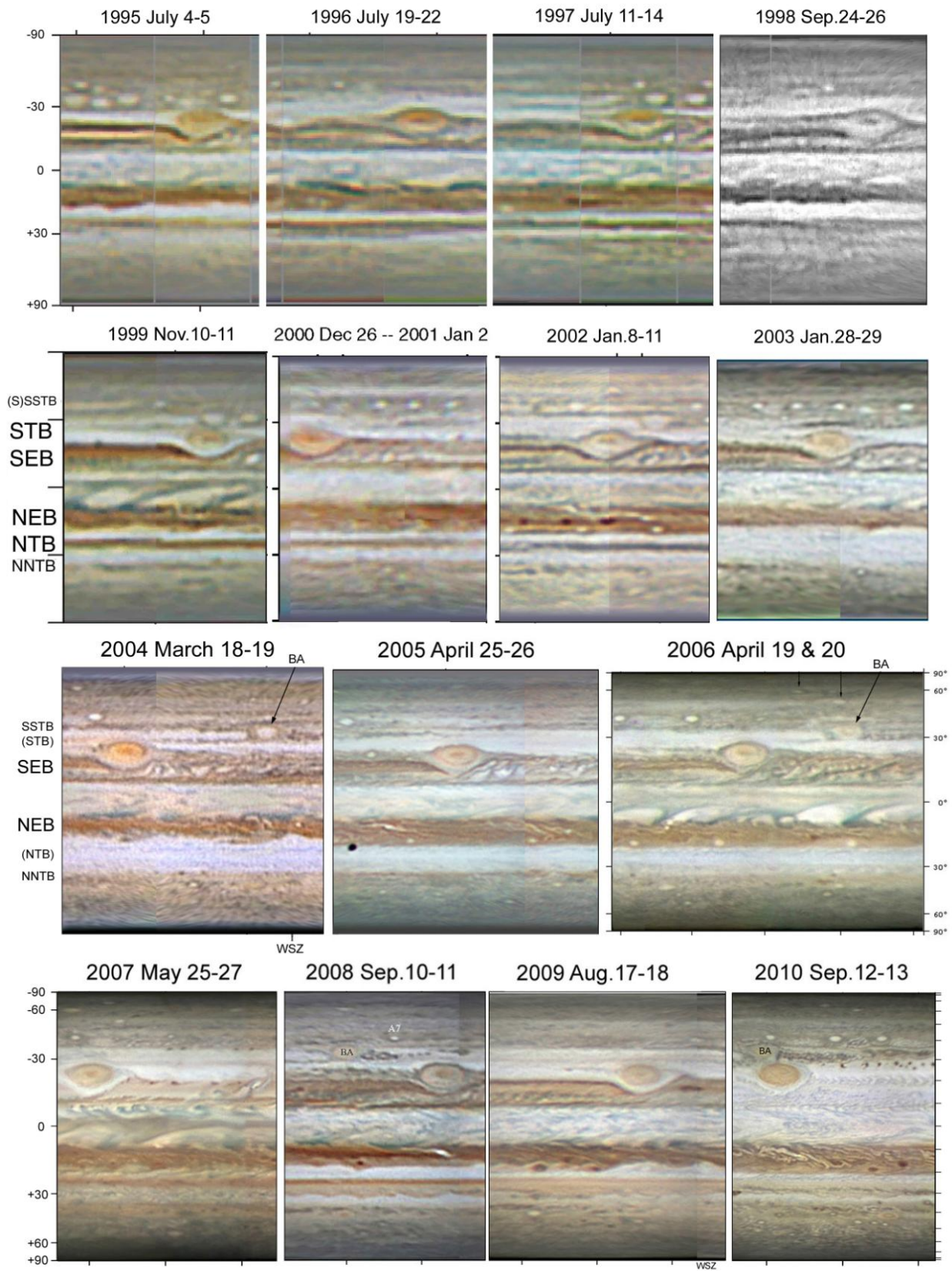

**Figure 2.**

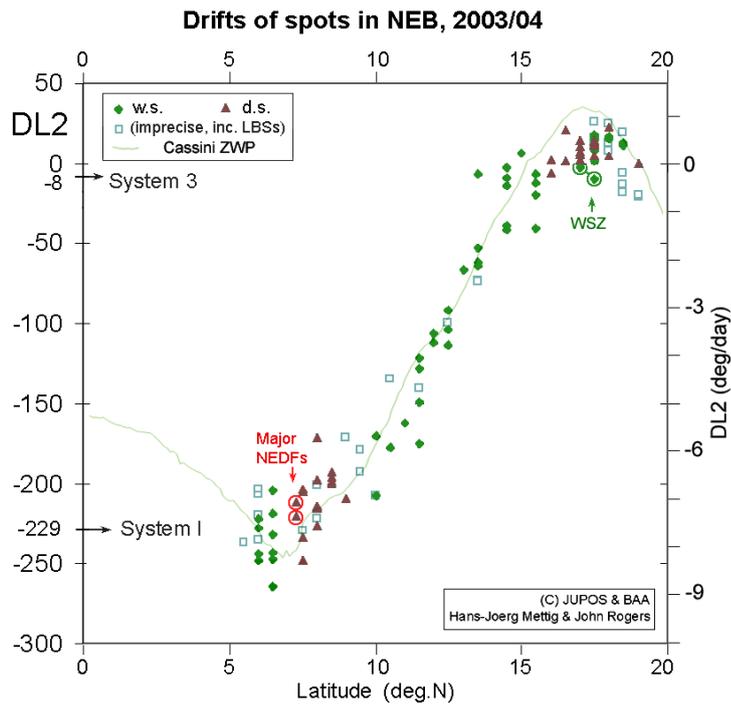

**Figure 3**(A). Chart of speed vs. latitude across NEB in 2003/04: [Ref.30, report no.5].
Latitudes of white spots (w.s.) and dark spots (d.s.) were assigned to the nearest half-degree by comparison of the visibility of tracks on JUPOS charts at one-degree latitude increments, prepared by H-J. Mettig. Most drift rates are accurate to ±4 deg/month, but open squares indicates less precise drift rates (due to spots with shorter lifetimes or more variable drifts), accurate to perhaps ±10 deg/month. In the range 17-19ºN, these open squares are sequential values for Little Brown Spots which were oscillating with periods ~20 days. Symbols are circled for large spots which would have been tracked visually: white spot Z on NEBn, and averages for two groups of dark projections on NEBs. The larger scatter of drift rates at low latitudes may be because some spots are assocaietd with the NEDFs, while others have faster speeds following the overall gradient.

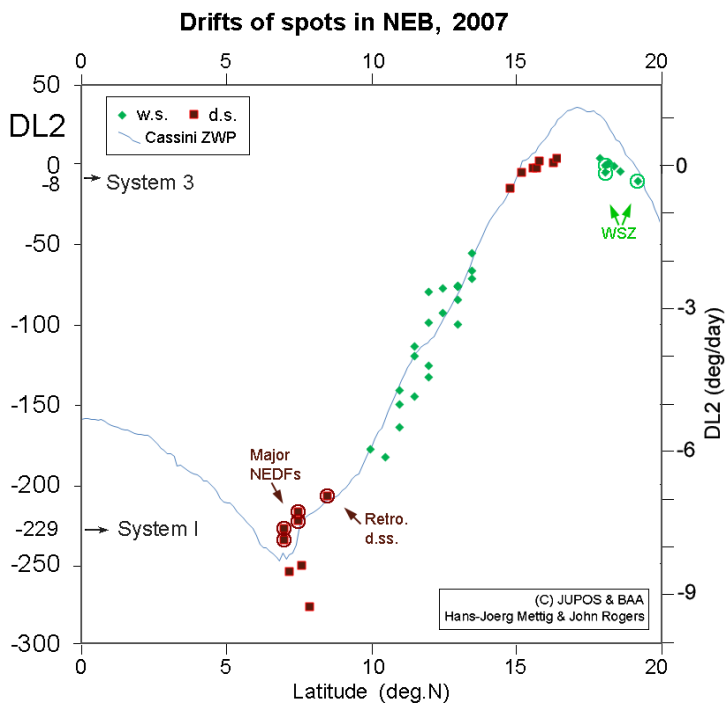

**Figure 3**(B). Chart of speed vs. latitude across NEB in 2007, produced as in (A); enlargement of data in our 2007 report [Ref.20].

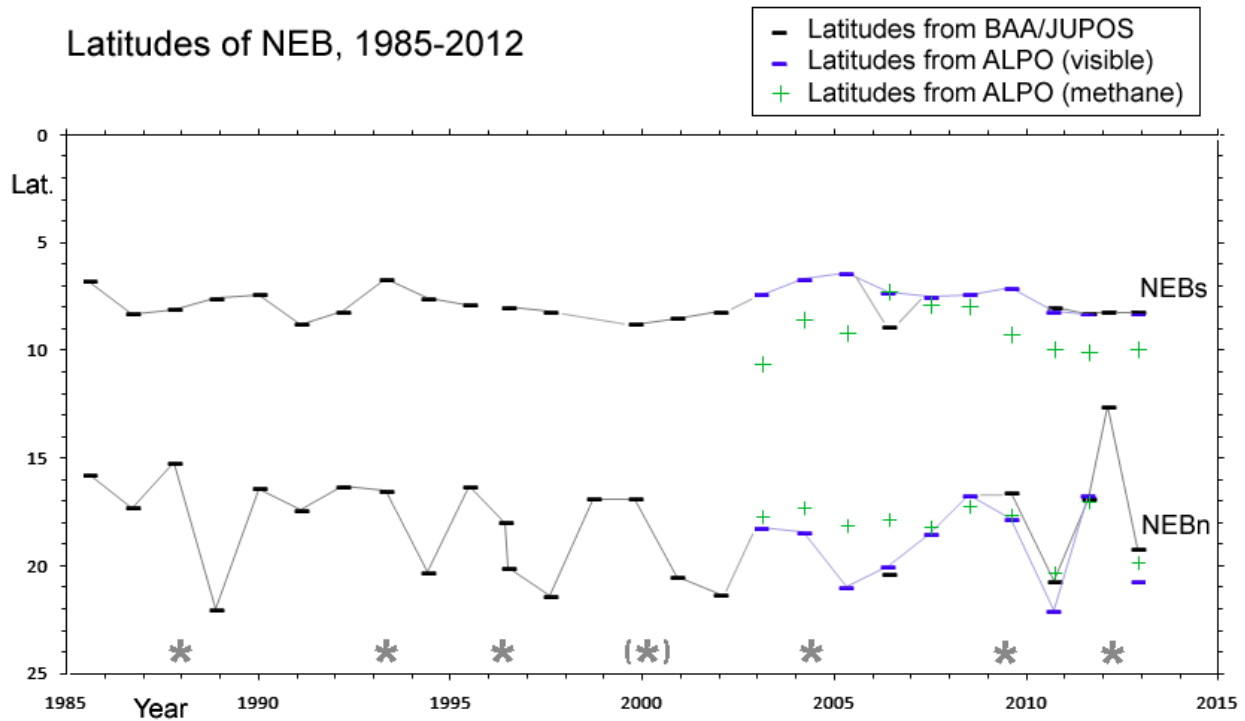

**Figure 4**. Chart of latitudes of the N and S edges of the NEB over the years 1985-2012.
Asterisks indicate times of onset of NEEs. Data are from the following sources: 1985-2002 (black), BAA/JUPOS apparition reports published in the Journal of the BAA (except 1993 and 1994 which are still unpublished); 2002-2012 (blue and green), ALPO apparition reports published by R. W. Schmude in the Journal of the ALPO; 2009-2012 (black), estimates from hi-res maps produced by Marco Vedovato of the JUPOS team. Green crosses are measurements from methane-band images (0.89 μm), in which the NEBs was often at much higher latitude than in visible images, and the NEBn sometimes did not change during visible-light NEEs.

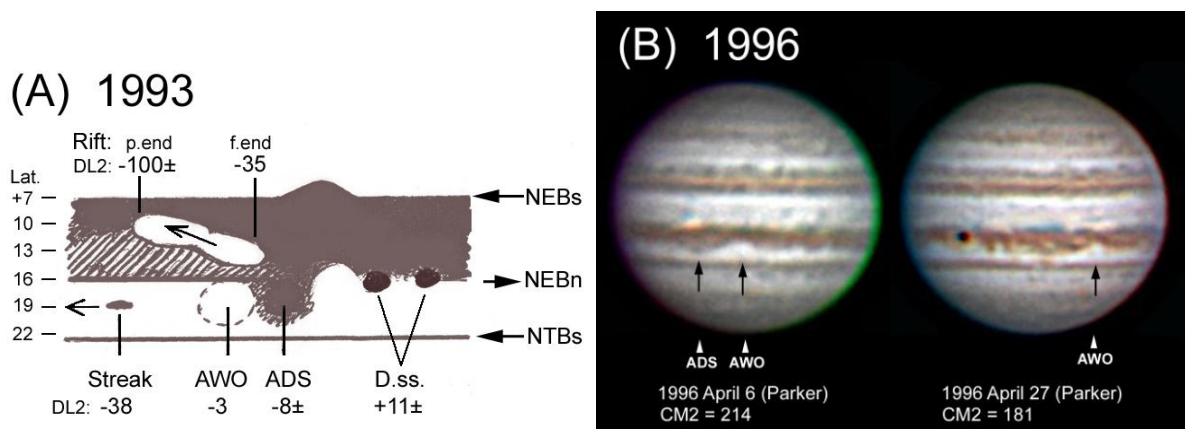

**Figure 5**. Images of early stages of four NEEs, including NEBOs. (A) 1993; (B) 1996; (C) 2004; (D) 2009.
(A) The NEBO at the origin of the NEE in late March, 1993, summarised from drawings and photographs. Numbers are drifts (DL2) in degrees per month.
(B) Origin of the NEE in 1996 April. On April 6 (the first image to show it) there is a newly formed ADS and bulge just p. an AWO. By April 27, a spotty expansion of NEBn has spread some way p. this AWO. (Colour fringing of the disk and satellite shadow are due to time intervals between the colour channels.)

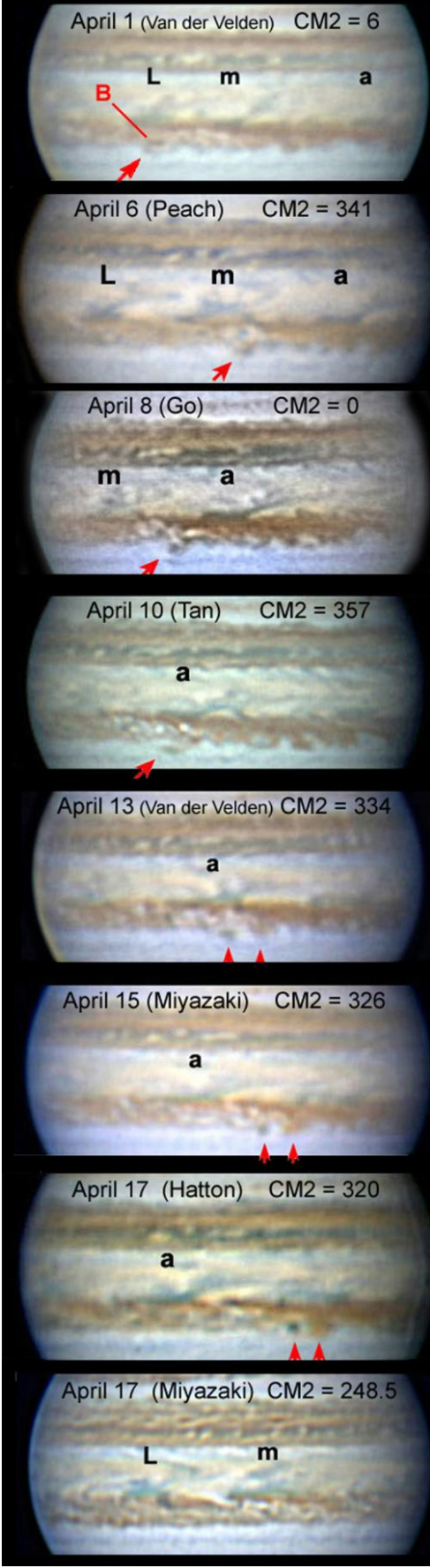
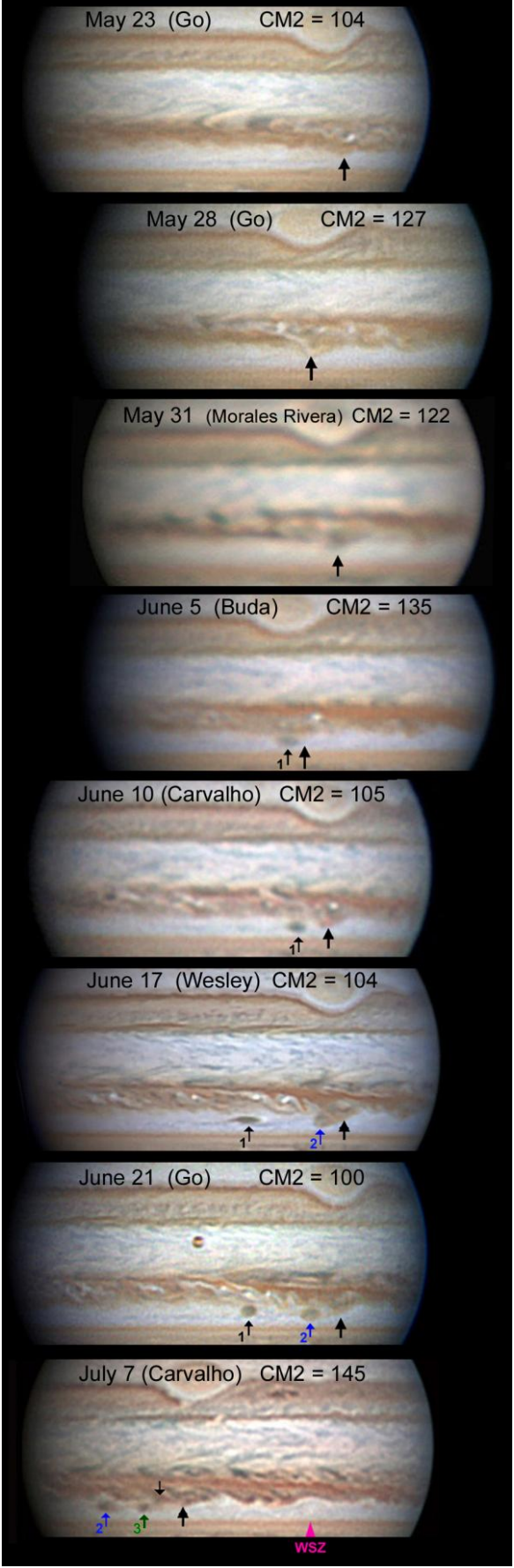

**Figure 5 [cont.]**. (C) Origin of the NEE in 2004 April. On April 1, a NEBO begins as an ADS (red arrow) appears adjacent to a pre-existing barge (B), while a large NEB rift system is passing it. New bright rifts then proliferate around this longitude. While the first ADS changes from grey to brown, a second, very dark grey ADS appears 10º p. it on April 13. Meanwhile, NEBs dark formations labelled L, m, and a drift past; m and a are disturbed as they pass the outbreak. The final panel shows the sector p. the NEBO, filled with the large pre-existing rift system.
(D) Origin of the NEE in 2009 May-June. On May 23 a brilliant white spot appears in a pre-existing rift system (arrow). By May 28, this develops into rift pushing northwards into the NTropZ, which generates a very dark grey ADS (May 31). This is the first of three ADSs which are formed at this site and prograde (numbered arrows).

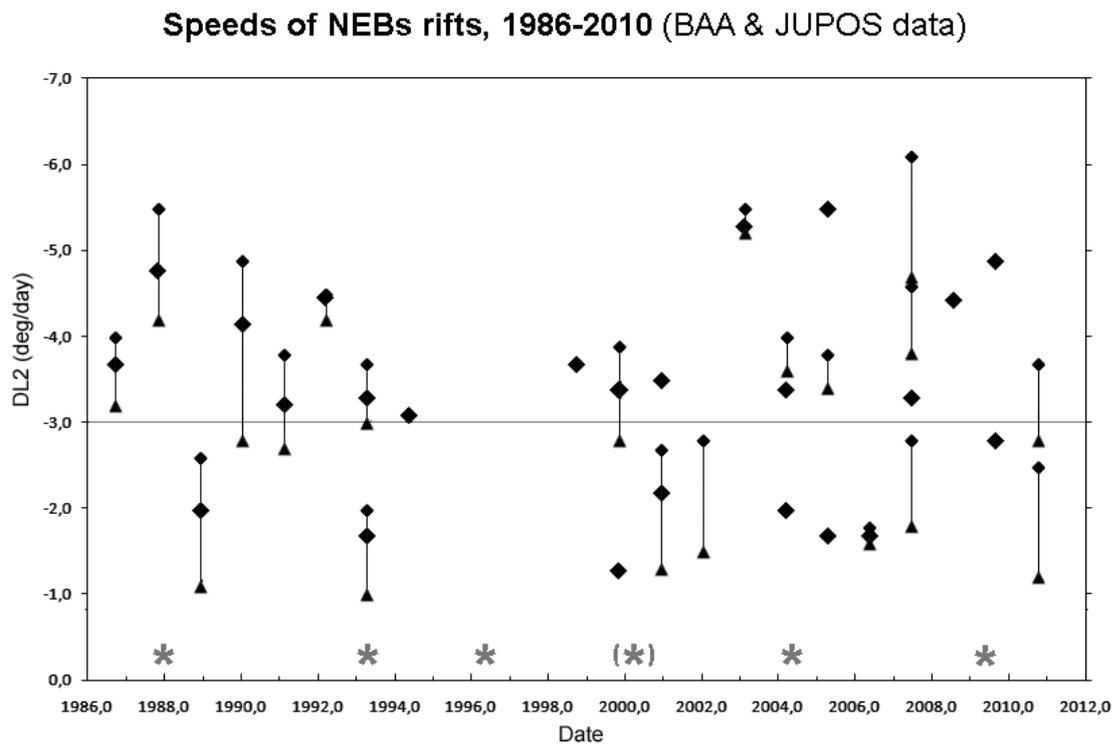

**Figure 6**. Chart of speeds measured for NEB rifts, 1986-2010 [Ref.29]. Large diamonds, averages or single points; small diamonds and triangles, range of observed speeds; asterisks, dates of onset of NEB expansion events. Note that speeds are largely segregated into ranges above or below -2.8 deg/day.

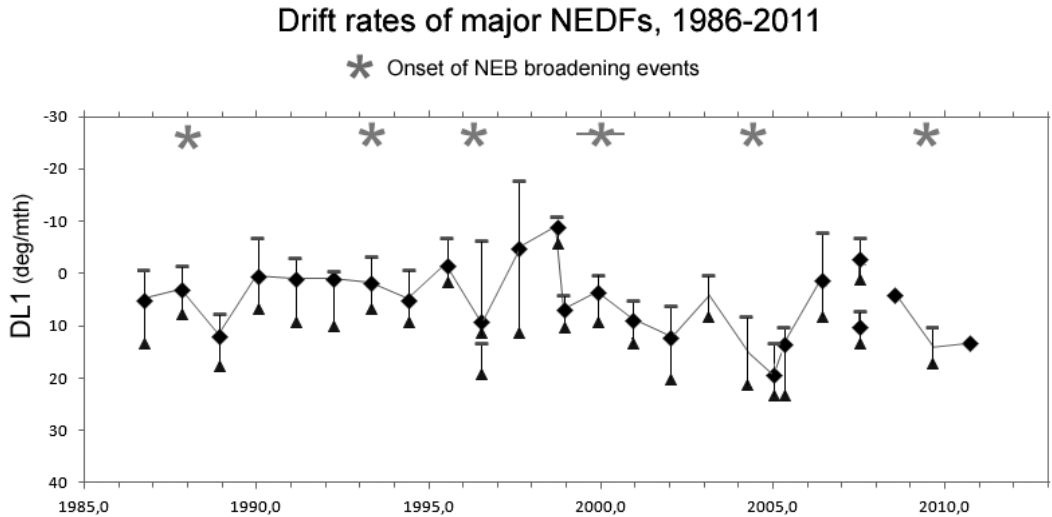

**Figure 7.** Chart of NEDF speeds in each apparition. 1986-2010. Bars show the range of speeds observed. Only the major NEDFs are included, so the list is comparable with the historical record and does not include small features which might have different characteristics. Asterisks indicate times of onset of NEEs.

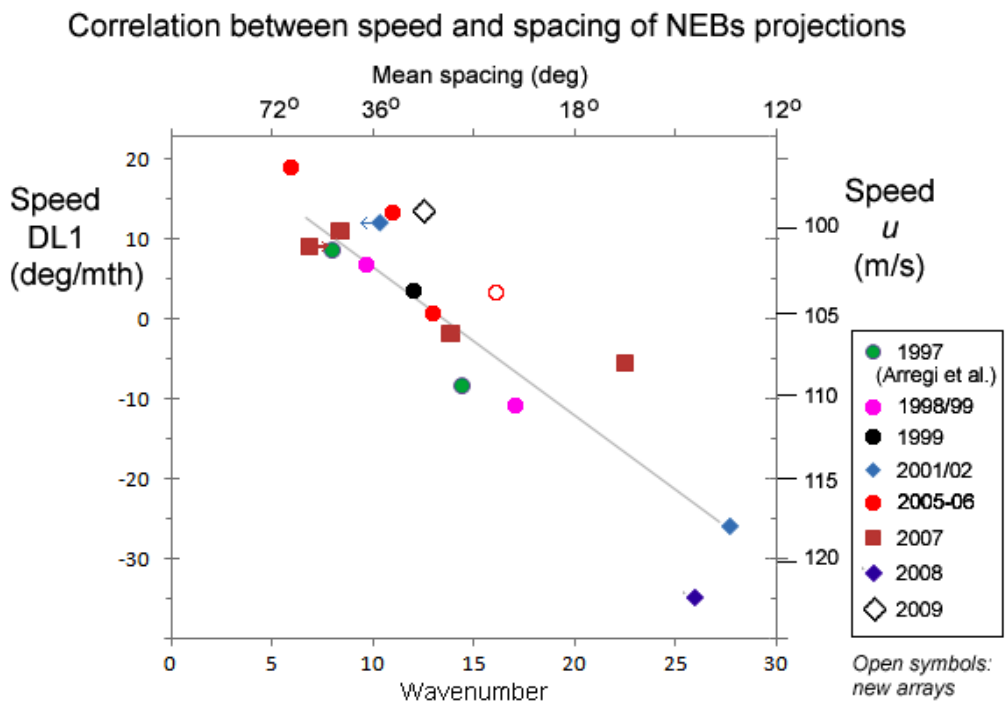

**Figure 8.** Correlation between speed and spacing for NEDFs, from BAA/JUPOS reports, plus 1997 points from [Ref.18]. Chart updated from [Ref.20]. Points are only plotted for times when there were arrays of NEDFs with reasonably uniform spacing. Open symbols represent poorly-fitting values, which were for newly-created arrays of NEDFs, suggesting that the relationship may be disturbed in times of transition. This chart includes 'fast' but not 'super-fast' features.

___________________________________________________________________

## *Appendices (Supplementary Online Material)*

*I.  JUPOS chart of NEBn/ NTropZ from our long-term WSZ report [below, from Ref.28:]*
    http://www.britastro.org/jupiter/2013_14report03.htm

*II.  Summaries of the observations year-by-year for NEB rifts [already posted, Ref.29:]*
    http://www.britastro.org/jupiter/relationnebrifts.htm

*III.  Ditto for NEDFs [below]*

*IV.  JUPOS/BAA chart of NEBs formations in 1998/99  [below].*
___________________________________________________________________

**APPENDIX I:
JUPOS chart for
N.Tropical domain,
2001-2013 (in L2).**
Track of WSZ is marked in red.
(From Ref.28)

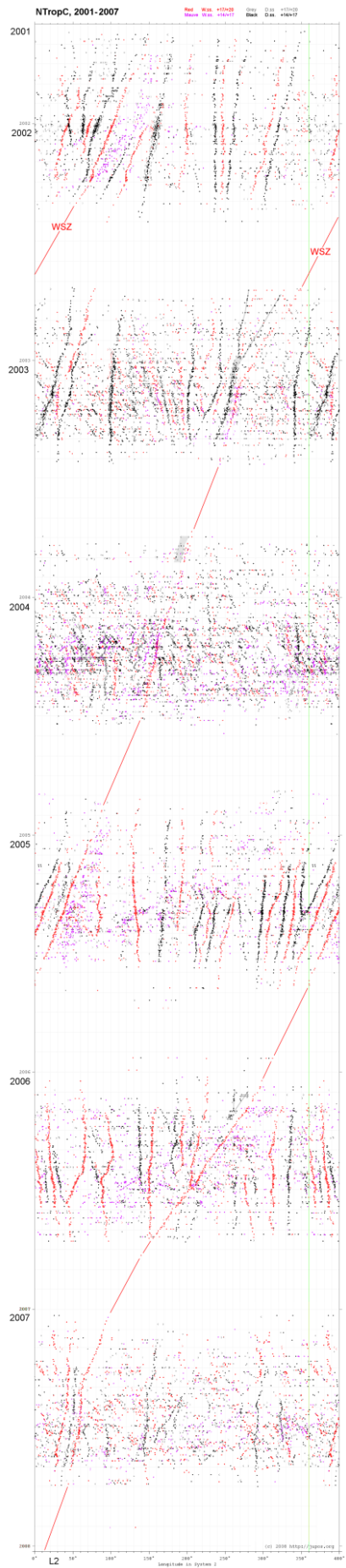
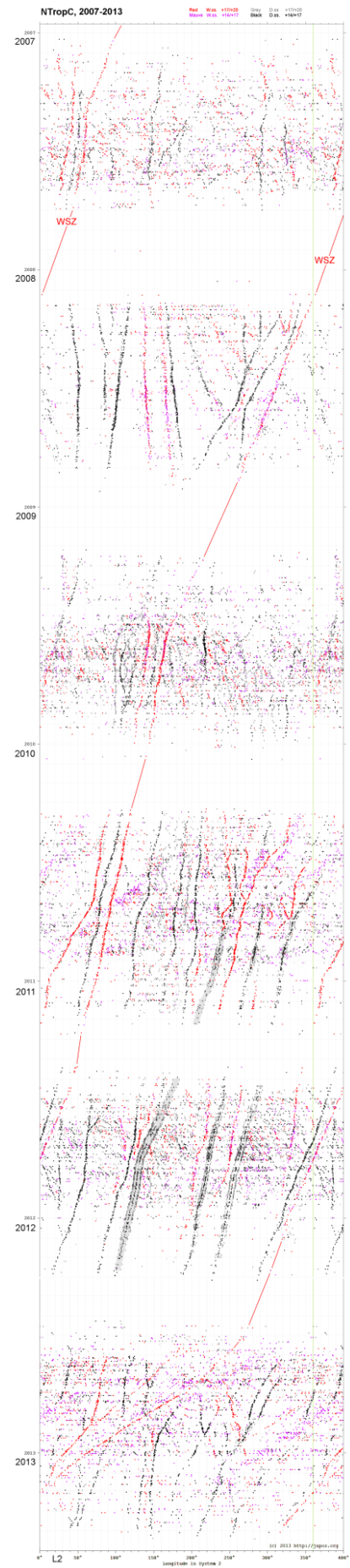

# APPENDIX III: Chronicle of appearance of the NEDFs, 1986-2010

*These notes summarise the appearance and behaviour of the NEDFs (including 'fast' features when present), in each apparition, from the same sources listed in the main text, viz: published BAA reports up to 2002, on-line BAA/JUPOS reports from 2005-2009, and unpublished data as noted below.*

```
In 1986/87, several large dark plateaux on NEBs 'collapsed' as rifts passed, then
disappeared, while their p. ends appeared to prograde rapidly.

In summer 1987, there were still some long plateaux but some largely featureless
sectors.  NEDFs were very variable, some large, some again 'collapsing' in the p.
direction, with rift interactions.  From 1987 Sep. onwards, more NEDFs were
developing.  (NEE started at the end of Dec.)

In 1988/89, NEBs features were inconspicuous, not projecting into EZ.

In 1989 Aug-Sep., NEBs remained quiet. In Oct-Nov. a great NEB rift induced an
impressive series of NEDFs, but they mostly subsided after ~2 weeks. From 1990 Jan.
onwards there were more substantial NEDFs (probably the earlier small ones were
revived by a rift).

In 1990/91 there was a complete array of NEDFs, spacing 25-35°, mostly large and
well-formed and stable.

In 1991/92 and 1993 there was still an array of major NEDFs, many being stable. In
1994 there were still some typical major NEDFs, though some sectors had only minor
features [BAA, unpubl.]. In 1995, NEDFs were still quite conspicuous but variable and
not stable for long.

In 1996 (after the NEE started in early April), there were typical dark plateaux all
round in late April, but by late May the NEBs was more broken up and variable.  Many
NEDFs became large and dark and bluish since the NEE. They included 4 large slow-
moving plateaux (DL1 = +13 to +19), but also a sector of 6 more variable projections.

In 1997, NEDFs were irregular and very rapidly varying, with diverse drifts.

In summer 1998, the NEBs was exceptionally disturbed, with 3 sectors: (i) NEDFs
subdued or chaotic; (ii) 4 NEDFs with DL1 ~ -11, spacing 21°; (iii) 4 NEDFs with DL1
~ +4 to +9, spacing 38°.  In autumn, drifts reversed and some oscillated.  By early
1999, all sectors had more coherent DL1 ~ 0 to +10.

In 1999/2000, there was an exceptionally conspicuous array of 12 projections with
well-formed dark festoons, leading into a very dark EB; drifts were fairly stable
(DL1 = 0 to +9).

In 2000/01, there were still very dark festoons and EB up to 2000 July, but these
were broken up in Aug-Sep., and some NEDFs disappeared, leaving only 7.  By Dec., the
festoons and EB had almost disappeared.

In 2001/02, there were only 4-5 remaining large features (long low plateaux with thin
festoon). Meanwhile there were also numerous fast-moving small spots, both bright
spots in EZ(N) and dark projections.

In 2002/03, there were plenty of large projections but they were variable and
transient; and many short-lived rapidly-moving spots, mostly bright spots in EZ(N)
but also some dark projections. [JUPOS, unpubl.]

In 2003/04, there were several examples of NEDFs altered by passing rifts. In early
2004, there was a well-defined array of 11 large dark projections, and again some
fast-moving small spots; these cut across the positions of some large features, but
were mostly seen at the former locations of two which were broken up by passing
rifts.  Again these fast features included many bright spots (6 to 6.5°N) and some
dark spots (7.5°N). (There were also many spots with exceptional positive DL1, at 8-
9°N.) [JUPOS, unpubl.]
```

In 2004/05 *[updated from final report]*: there were initially 6 long plateaux with extreme positive DL1; disrupted by rifts in 2005 Feb. and replaced by a new array of NEDFs. (All at ~8 deg.N.)

In 2006 *[updated from final report]*, there was a major change, last seen in 1999: the NEBs was dominated by a very regular array of 12 formations with faster speed (DL1 ~ +1), whose festoons were much darker than before, as was the EB; this led up to the global upheaval of 2007. (A pair of very fast white spots, DL1= -27, at 8.2°N, in plume cores, were atypical. There were also small retrograding features.)

In 2007, The dark projections and festoons were still very prominent, dark blue-grey, linked to the very dark brown Equatorial Band.  All these features gradually faded to moderate darkness during the summer. There were about 12 well-defined projections, but they were quite variable in appearance and drifts, with a number of mergers, splittings, and new appearances during the year.  There appeared to be two concurrent sets – one set of compact projections with zero or slightly negative DL1, and a set of longer, more widely spaced plateaux with positive DL1 – probably superimposed as wave-trains showing interference. Two of the formations consisted of successive fast-moving dark projs., DL1 from -17 to -37 (at 7.3°N).
There was also one white spot (7.9°N) with DL1 = -47 for one week.

In 2008, the typical large projections were subdued and few in number, and the space vacated by them was occupied by small spots and projections with unprecedentedly fast speeds ***[see main text]***.

In 2009, up to June, there were no large features on NEBs, but there were minor transient festoons with DL1 ~ -18 to -40 (mean 7.0°N).  But in July, all was transformed with the appearance of many retrograding dark spots, large and small, probably generated by the adjacent rift system associated with the ongoing NEE. There were 10 of them by Sep-Oct., with a typical spacing of 29°.  Only a few fast features were detectable up to early Sep., and none thereafter.

In 2010 and 2011/12:  *See Paper II [new analysis for this report].*

____________________________________________________________________

# APPENDIX IV.   JUPOS/BAA chart of NEBs formations in 1998/99.
JUPOS chart of all visual and photographic measurements (produce by H-J. Mettig, annotated by JHR).

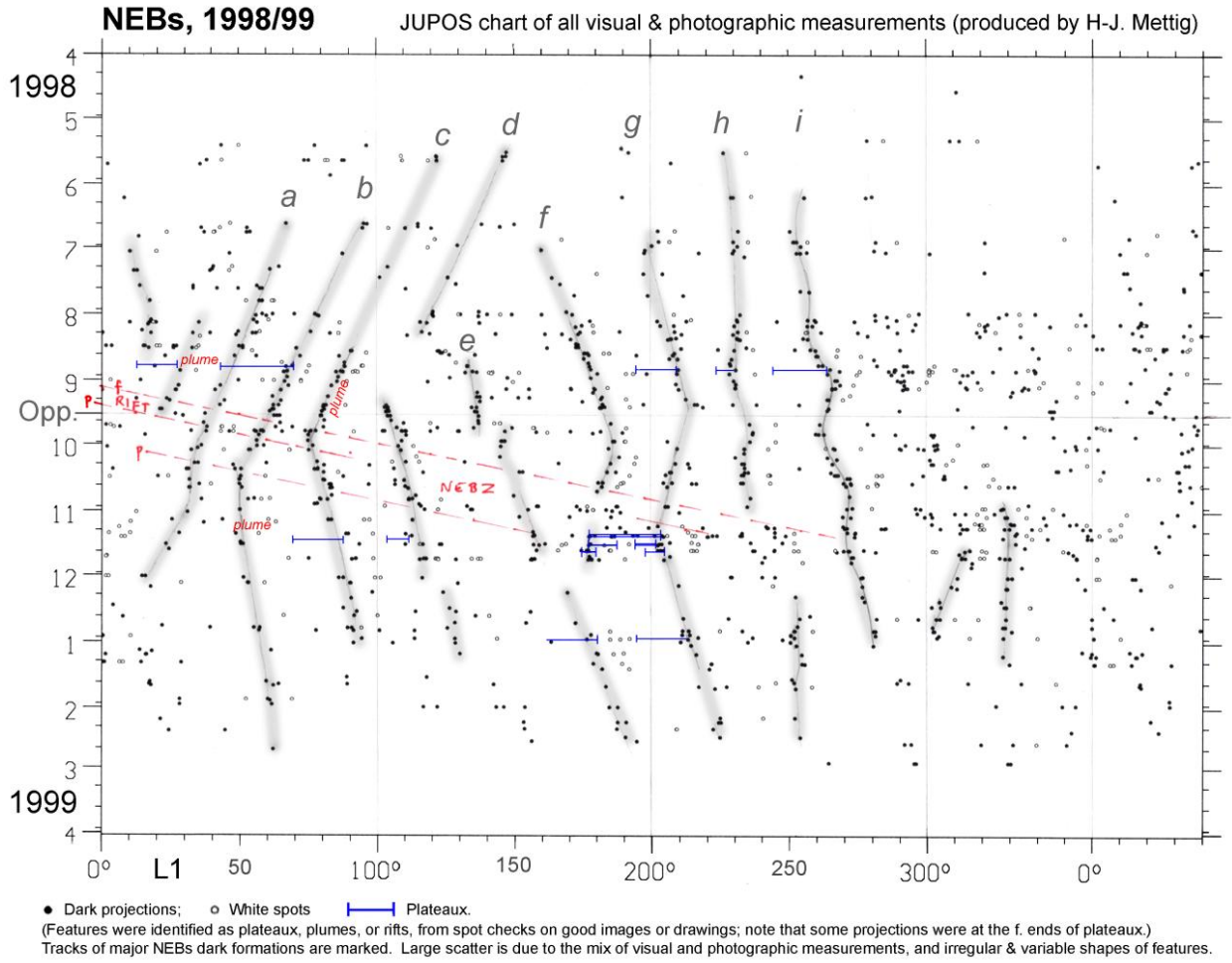